\documentstyle[aps,twocolumn]{revtex}

\begin{document}


\title{Black hole entropy in the Chern-Simons formulation of $2+1$
gravity}

\author{M\'aximo Ba\~nados$^{1,2}$ and Andr\'es Gomberoff$^{1}$ }

\address{$^1$Centro de Estudios
Cient\'{\i}ficos de Santiago, Casilla 16443, Santiago, Chile\\
$^2$Departamento de  F\'{\i}sica, Universidad de Santiago,
Casilla 307, Correo 2, Santiago, Chile}

\maketitle

\begin{abstract}

We examine Carlip's derivation of the 2+1 Minkowskian black hole
entropy. A simplified derivation of the boundary action --valid for
any value of the level $k$-- is given. 
\end{abstract}

\section{Introduction}

In the last two years there has been major progress in the
understanding of the quantum mechanics of black holes. On the one
hand, Carlip \cite{Carlip} has given a statistical description for
the entropy of the 2+1 black hole\cite{BTZ}. More recently,
string theory has also provided a statistical description of the
black hole entropy for extremal and near extremal black holes
\cite{Strings}. Despite the success of these new formulations
much work remains to be done. In fact, Carlip's approach relies
heavily on the Chern-Simons formulation of 2+1 gravity and therefore
its generalization to higher dimensions is not an easy
task\cite{gegenberg}.  The formulation given in \cite{Strings}, on
the other hand, can be implemented in various dimensions but only for
extremal and near extremal black holes. The real 4-dimensional
non-extremal black hole still seems far from being completely
understood.   

In this note we address some issues concerning Carlip's 
derivation for the entropy of the Minkowskian 2+1 black hole.  It was
shown in \cite{Carlip} that the degeneracy of boundary degrees
of freedom of 2+1 gravity gives the correct value for the black hole
entropy. However, the explicit form of the boundary action was not
written in \cite{Carlip} because it involved a complicated Jacobian.
It was argued instead that in the limit $k\rightarrow \infty$ the
boundary degrees of freedom should be described by Kac-Moody currents
subject to the constraint $L_0=0$ (this constraint was imposed
because $L_0$ generates a gauge symmetry at the boundary). Here we
shall prove that the Kac-Moody currents are indeed the relevant
degrees of freedom for any value of $k$, and the constraint $L_0=0$
is also necessary to ensure differentiability of the three
dimensional  action.  We also find the explicit formula for the WZW
action that gives rise to the boundary degrees of freedom for any
value of $k$.  Our analysis is simple and relies only on some general
considerations of Chern-Simons theory formulated in a manifold with a
boundary. However, the quantization of the resulting boundary theory
(which is classically well-defined for all values of $k$) will be
possible only in the limit $k\rightarrow\infty$. The reason is that
the WZW action for the group $SO(2,1)$ is not completely understood.
In particular, we do not know how to count states in the full
non-Abelian theory. 

Carlip's analysis has two main ingredients. First, it is assumed
that the entropy can be associated to a field theory lying at the
horizon. This assumption has been extensively discussed in the last
few years by Carlip himself \cite{Carlip2} and others
\cite{Balachandran2,Maggiore}. One can further justify it by
resorting to the $0^{th}$-law of black hole mechanics
which states that the surface gravity $\kappa$ is constant over the
horizon.  Therefore, the thermodynamic object $is$ the horizon and it
is thus natural to look for microscopic states defined on that
surface.  Second, the horizon is assumed to rotate with a rigid
angular velocity and that parameter -which only depends on time- is
varied in the boundary action principle. Given these two
assumptions the rest is done by the dynamics of 2+1 gravity. It only
remains to set appropriate boundary conditions to ensure the
existence of a black hole, find the boundary action and quantize
it\footnote{As stressed in \cite{Carlip}, due to the non-compact
nature of the symmetry group and the lack of a full diffeomorphism
invariance, the resulting Hilbert space has states with negative
norm.}. 

In this paper we shall mainly be concerned with the issue of imposing
the correct boundary conditions and finding the boundary action; we
shall not attempt to clarify or further analyze the two assumptions
described above. As we shall see, the method followed here to find
the boundary action is remarkably simple and may be, in principle,
applicable to 3+1 dimensions. 

For notational simplicity and to gain some generality we shall start
by analyzing the problem of boundary conditions in Chern-Simons
theory for a general Lie group $G$. Once the general case is
understood the application to 2+1 gravity will be straightforward.    

\section{Classical Chern-Simons theory on a manifold with a boundary}

\subsection{The action}

In this section we introduce some general aspects of Chern-Simons
theory on a manifold with a boundary. We consider a Chern-Simons
action formulated on a manifold $M$ with the topology  
$\Sigma\times\Re$ and the ``spacelike" surface $\Sigma$ has the
topology of an annulus. The manifold $M$ has thus two disconnected
``timelike" boundaries given by 
\begin{equation}
B_+ = \partial\Sigma_+ \times \Re, \mbox{\hspace{1cm}} B_\infty=
\partial\Sigma_\infty \times
\Re 
\end{equation}
where $\partial\Sigma_+$ and $\partial\Sigma_\infty$ are the
boundaries of $\Sigma$.  Since both $\partial \Sigma_+$ and $\partial
\Sigma_\infty$ are topologically circles, $B_+$ and $B_\infty$ are
cylinders.   

An important difference between the inner ($B_+$) and outer
($B_\infty$) boundaries is that $B_\infty$ is located at an  infinite
distance while $B_+$ is located at a fixed finite distance. As it has
been proved in \cite{Brown-Henneaux2}, the asymptotic group (the
group of transformations that leave the asymptotic conditions
invariant) at $B_\infty$ has a classical central charge. This central
charge is absent at the inner boundary because $B_+$ is located at a
finite distance and therefore diffeomorphisms normal to the boundary
--responsible for the central charge-- are not accepted \cite{B}. 

The Chern-Simons action is given by 
\begin{equation} 
I_{CS} = k W[A] + B 
\label{I} 
\end{equation} 
where
\begin{equation}
W[A] = \frac{1}{4\pi}\int_M Tr(AdA + \frac{2}{3}
A^3) 
\label{W}
\end{equation}
is the Chern--Simons functional, and $B$ is a boundary term. Its
variation gives rise to the equations of motion $F =0$, where $F= dA
+ A\wedge A$ is the Yang-Mills curvature 2-form . These equations can
be
split in the convenient 2+1 form 
\begin{eqnarray} 
\dot A^a_i &=& D_i A^a_t,
\label{Ad}\\
F^a_{ij}&=&0 
\label{c} 
\end{eqnarray} 
showing that the time evolution is generated by a gauge
transformation with parameter $A^a_t$.  Eq. (\ref{c}) is a
constraint over  the initial conditions.  Here we have denoted by
$x^0=t$ the coordinate running along $\Re$, and $x^i$ are local
coordinates on $\Sigma$.  

An important point to ensure the validity of the above equations 
is the cancellation of the boundary term
\begin{equation}
-\frac{k}{4\pi}\int_{\partial M} Tr\ A\wedge \delta A + \delta B=0
\label{B}
\end{equation}
which appears when (\ref{I}) is varied. As usual,  at the initial and
final boundaries (\ref{B}) is canceled by imposing $\delta A=0$ and
$B=0$. However, in our present case there are two other timelike
boundaries, namely $B_+$ and $B_\infty$.  The treatment of the outer
boundary $(B_\infty)$ is
standard and we shall not repeat it here. The interested reader can
consult \cite{Brown-Henneaux2,Coussaert-Henneaux-vanDriel,Ezawa} for
the case of gravity and \cite{Bimonti,B} for the general case. We
will concentrate here in the inner boundary which in the next section
will be associated to the black hole horizon.   
 
\subsection{Boundary conditions}

Let $\varphi$ be an angular coordinate running along
$\partial\Sigma_+$
and  $x^0=t$, then the boundary term (\ref{B}) at the inner 
boundary reads 
\begin{equation}
-\frac{k}{4\pi} 
\int_{B_+} dtd\varphi\ Tr(A_t \delta A_\varphi - A_\varphi
\delta A_t) +\delta B =0. 
\label{B2}
\end{equation}
A simple way to cancel (\ref{B2}) is by imposing the boundary
condition $A_t=0$ and $B=0$. The group of gauge transformations
leaving these boundary conditions invariant are those 
whose parameters do not depend on time. These transformations are
global symmetries and are generated by Kac-Moody
currents\cite{Witten89,Moore-Seiberg}. A second possibility to ensure
the vanishing of (\ref{B2})  is to set $A^a_t$ equal to a fixed given
value, i.e., $\delta A^a_t=0$ at $B_+$. We then set $B= (k/4\pi) \int
Tr(A_t A_\varphi)$ producing an action which has well defined
variations. The residual group in this case is given by the set of
parameters $\lambda^a$  satisfying $D_0
\lambda^a=\dot\lambda^a+[A_t,\lambda]^a=0$. Thus, in this case the
parameters can depend on time but their dependence is  not arbitrary
because $A_t$ is fixed. Again, these transformations are generated by
Kac-Moody currents and they are global transformations.    
   
In our application to black hole physics, we will need a different
set of boundary conditions.  Consider the case on which the surface
$\partial \Sigma_+$ (which is topologically a circle) rotates
with
angular velocity $w(t)$. Since the time evolution is generated by a
gauge transformation with parameter $A^a_t$ [see Eq. (\ref{Ad})], the 
appropriate boundary condition is,   
\begin{equation} 
A_t = w(t) A_\varphi
\label{A0}  
\end{equation}
because, in Chern-Simons theory, a displacement in $\varphi$
with parameter $w(t)$ is equivalent to a gauge transformation with
parameter $w(t)A^a_\varphi$ \cite{Witten88}.

Having chosen the boundary conditions we now have to address two
remaining things. First, whether the boundary conditions (\ref{A0})
are enough to ensure the differentiability of the action. Second,
what is the set of gauge transformations that leave (\ref{A0})
invariant. These two issues are connected.   

Under (\ref{A0}) the boundary term (\ref{B2}) reduces to 
\begin{equation}
\frac{k}{4\pi}\int_{B_+} dtd\varphi\ Tr(A_\varphi^2) \delta w(t)
+\delta B=0.
\label{B3}
\end{equation}
To ensure the vanishing of this boundary term we have
two possibilities. One could impose $\delta w(t) =0$ and $B=0$. In
this case, the surface rotates with a given --fixed-- angular
velocity.  A second possibility -which will be the relevant boundary
condition for the black hole-- is to vary with respect to $w(t)$.
This implies that the coefficient of $\delta w(t)$ in (\ref{B3}) must
vanish, which in turn ensures the differentiability of the action
(with $B$=0).

Indeed, if $w(t)$ is varied there exists a
$gauge$ symmetry at the boundary whose generator is the coefficient
of $\delta w(t)$ in (\ref{B3}). This can be seen as follows.  We
look at the most general set of gauge transformations $\delta A^a_\mu
= -D_\mu \lambda^a$ leaving (\ref{A0}) invariant. This group
will be called `the boundary group' at $B_+$.   One finds
the condition over $\lambda^a$, 
\begin{equation}
\dot \lambda^a = -\delta w(t) A^a_\varphi +
 w(t) \partial_\varphi \lambda^a.
\label{res}
\end{equation}
Note that, since $w(t)$ is not fixed, we have allowed for
transformations with $\delta w\neq 0$. 

The boundary group has two pieces. First, for those
transformations with $\delta w(t)=0$ one finds that the time
derivative of $\lambda^a$ is completely determined by (\ref{res}).
These are global symmetries and are generated by Kac-Moody
currents. A different solution to (\ref{res}) is provided by     
\begin{equation}
\delta w(t) = -\dot \epsilon(t), \mbox{\hspace{1cm}} 
\lambda^a = \epsilon(t) A^a_\varphi, 
\label{res2}
\end{equation}
where $\epsilon(t)$ is an $arbitrary$ function of time and
$A_\varphi$ satisfies its equation of motion.   This is a gauge
symmetry because it contains an arbitrary function of time. The
transformation (\ref{res2}) corresponds to rigid
($\varphi$-independent) time-dependent rotations of the surface
$\partial\Sigma_+$ \cite{Carlip}. The generator of these rotations is
the
zero mode of $g_{ab}A^a_\varphi A^b_\varphi$ which should then vanish
because its associated transformation is a gauge symmetry.  Going
back to (\ref{B3}) we see that the vanishing of 
\begin{equation}
L_0 \equiv \frac{k}{2} g_{ab} A^a_\varphi A^b_\varphi |_{zero\ mode}=
0
\label{L0=0}
\end{equation} 
also ensures the differentiability of the action (with $B=0$). In
summary, the group of transformations that leave the boundary
conditions (\ref{A0}) invariant is given by the semidirect product of
the Kac-Moody symmetry times the (time-dependent) rigid translations
along $\varphi$. Note that $L_0$ is the zero mode Virasoro operator
of the theory.  [Only the zero mode Virasoro constraint appears
because $w(t)$ does not depend on $\varphi$.]

\subsection{The induced theory at the boundary}

Having chosen the boundary conditions we can now study the induced
theory at the boundary. As it is well known, Chern-Simons theory
in 2+1 dimensions does not possess local degrees of
freedom\footnote{It has been proved in \cite{BGH} that this property
is not carried over to higher dimensional Chern-Simons theories. For
$D>3$, the gauge symmetries are not enough to kill all the degrees of
freedom and local excitations do exist.} so fixing the gauge will
leave us only with some global degrees of freedom. These global
degrees of freedom can be of two types. On the one hand, there may be
non-trivial holonomies. This is certainly our case because the
spatial manifold has the topology of an annulus.  Another set of
degrees of freedom are the boundary values of the gauge field which
cannot be set equal to zero by an allowed gauge transformation. The
number of these states is infinite and for a fixed value of the black
hole area Carlip has shown that their degeneracy gives rise to the
correct value for the 2+1 black hole entropy \cite{Carlip}.    
 
Let us thus fix the gauge in order to isolate the boundary degrees of
freedom. As it is well known, the theory at the boundary is described
by a WZW model \cite{Witten89,Moore-Seiberg}. However, it is
instructive to obtain it directly from the equations of motion
projected to the boundary. An appropriate\cite{Miguel} gauge fixing
condition is \footnote{Here we fix the gauge in the interior. The
residual gauge freedom of the boundary conditions (\ref{res2}) is not
fixed by this gauge condition.}
\begin{equation}
A^a_r=0.
\label{A_r=0}
\end{equation}  
This gauge fixing condition together with the constraints (\ref{c})
simply imply that the tangential component of the connection,
$A^a_\varphi$, does not depend on the radial component. Thus,
hereafter we define
\begin{equation}
A^a_\varphi(t,r,\varphi) = A^a(t,\varphi) .
\end{equation} 
Eqs. (\ref{Ad}), on the other hand, contains the dynamical
information. The radial component, together with the gauge condition
(\ref{A_r=0}) allows the Lagrange multiplier $A^a_t$ to be solved. We
find that $A^a_t$ does not depend on $r$, which is also consistent
with the boundary condition (\ref{A0}). The angular component of
(\ref{Ad}) gives the dynamics of $A^a_\varphi$. Projecting to the
boundary and using (\ref{A0}) it reads,
\begin{equation}
\frac{d}{dt} A^a = w(t)\partial_{\varphi} A^a.
\label{eq}
\end{equation}
This equation together with the constraint (\ref{L0=0}) define
the dynamics at the boundary.  The values of $A^a$ at $B_+$ 
cannot be set equal to zero by an allowed gauge transformation. 

Equation (\ref{eq}) has the symmetries of the boundary conditions. 
Indeed, (\ref{eq}) is invariant under the gauge transformation
\begin{equation}
\delta A^a = \epsilon(t)\partial_{\varphi} A^a, \mbox{\hspace{1cm}}
\delta w(t) =
-\dot\epsilon(t) 
\label{gauge-sym}
\end{equation}
where $\epsilon(t)$ is an arbitrary function of $t$.  As stressed
above the generator of this gauge transformation is the zero mode
Virasoro constraint $L_0=0$ defined in (\ref{L0=0}). Eq. (\ref{eq})
has also the Kac-Moody global symmetry  given by the transformation
\begin{equation} 
\delta A^a = \partial_\varphi\lambda^a + [A,\lambda]^a,
\mbox{\hspace{1cm}} \delta w(t)=0
\label{sym3}
\end{equation}
where $\lambda$ satisfies the equation $\dot
\lambda=w\partial_\varphi \lambda$ [see Eq. (\ref{res})] but is
otherwise
arbitrary. Finally, (\ref{eq}) has also a global symmetry given by
the translations 
\begin{equation}
A^a \rightarrow A^a + \alpha^a, 
\end{equation}
where $\alpha^a$ is a constant Lie-algebra valued element. The
conserved quantities associated to this symmetry are the zero modes
of $A^a(\varphi)$ as can be directly verified from the equation
(\ref{eq}).

Eq. (\ref{eq}) is already in Hamiltonian form. We define
the (non-canonical) Poisson brackets,
\begin{equation}
[A^a(\varphi),A^b(\varphi')] = \frac{2\pi}{k}[ f^{ab}_{\ \ c}
A^c(\varphi) + k g^{ab} \partial_\varphi] \delta(\varphi,\varphi') 
\label{Poisson}
\end{equation}
and it is straightforward to check that (\ref{eq}) can be written in
the form
\begin{equation}
\frac{d A(\varphi)}{dt}^a = [A^a(\varphi),H], 
\end{equation}
where the Hamiltonian is 
\begin{equation}
H = \frac{k}{4\pi}\int d\varphi\ w(t) A^2.  
\label{H}
\end{equation}

The symmetries of (\ref{eq}) can also be written in Hamiltonian form.
The generator of the gauge symmetry (\ref{gauge-sym}) is the
Hamiltonian itself, while the generator of the Kac-Moody symmetry
(\ref{sym3}) is $K(\lambda) = \int \lambda^a A^b g_{ab}$. Note that
$K$ is a conserved quantity, and thus
a symmetry, only when $\lambda^a$ belongs to the
boundary group, that is, it satisfies $\dot\lambda^a = w
\partial_\varphi
\lambda^a$. 

We can now make contact with the well know fact that the dynamics at
the boundary of a Chern-Simons theory is described by a WZW model
\cite{Witten89,Moore-Seiberg}. Making the usual change of variables
$A=U^{-1} dU$ the above equations of motion can be derived from the
$1+1$ action
\begin{equation}
I = I_{\mbox{\tiny WZW}}(U) + \frac{1}{2\pi} \int dt d\varphi\  w(t)
L_0.
\label{action}
\end{equation}
Note that $w(t)$ enters in the action as a Lagrange multiplier. 
This action can thus be interpreted as a constrained WZW model in
which the variation of $I$ with respect to $w(t)$ imposes the
constraint $L_{0}=0$ among the Kac-Moody fields. 

The reader may notice that we have somehow re-derived the well known
relation between the WZW action and  Chern-Simons theory.  We have
chosen not to start with the WZW action from the very beginning to
stress the fact that, in principle, the method followed here could be
applied to 3+1 gravity.  The boundary theory can be found solely from
the boundary conditions and the equations of motion. The real problem
is the quantization of the resulting theory. The simplicity of the
2+1 theory relies in the fact that the quantization of a WZW model is
well understood for compact groups and that there are no bulk degrees
of freedom.  This allowed us to isolate the boundary degrees of
freedom in a simple way. 
 
The quantization of the above action is straightforward. The
canonical commutation relations (\ref{Poisson}) can be promoted to
quantum commutators without any trouble. The Hamiltonian $H$ is more
delicate because it has to be regularized. Fortunately, this problem
has been extensively studied in the literature. The correct quantum
Virasoro operator is 
\begin{equation}
L_0 = \frac{1}{2k+\hbar q} (T^2_0 + \sum_{n=1}^{\infty} T^a_{-n}
T^b_n g_{ab} ),
\label{qL01}
\end{equation}
where $q$ is the second Casimir in the adjoint representation and
the $T^{a}_{n}$'s are the Fourier components of $A^a(\varphi)$,
\begin{equation}
A^a(\varphi) =\frac{1}{k} \sum_{n} T^a_n e^{in\varphi}.
\label{A-modes}
\end{equation}
The commutator between the Fourier components $T^{a}_{n}$ is,
\begin{equation}
[T^{a}_{n},T^{b}_{m}]= if^{ab}_{c} T^{c}_{n+m} + k
n g^{ab}\delta_{n+m}
\label{TPB}
\end{equation}
and the normal ordered Virasoro operator satisfies the commutation
relation
\begin{equation}
[L_0,T^a_n] = -n T^a_n. 
\end{equation}
This commutation relation implies that $L_0$ has the form $L_0 =
C/(2k+\hbar q) + N$ where
\begin{equation}
C= g_{ab} T^a_0 T^b_0
\end{equation}
and $N$ is the number operator.  

\section{2+1 gravity and black hole entropy}

In this section we shall apply the results of the last section to the
special case of 2+1 gravity. As we shall see, this leads directly
to Carlip's formulation of the 2+1 black hole entropy\cite{Carlip}.  
The Chern-Simons formulation of 2+1 gravity consist on the sum of two
copies of the Chern-Simons action for the group $SO(2,1)$
\cite{Achucarro,Witten88},
\begin{equation}
I=kW[A] - kW[\tilde{A}] + B,
\end{equation}
where the Chern-Simons functional $W$ was defined in (\ref{W}).
The connections are related to the triad and spin connection through
\begin{equation}
A^a_\mu = w^a_\mu + \frac{e^a_\mu}{l}, \mbox{\hspace{1cm}} 
\tilde A^a_\mu =  w^a_\mu - \frac{e^a_\mu}{l},
\label{AA}
\end{equation}
where $A^a$ and $\tilde A^a$ are both $SO(2,1)$ connections and $l$
is a parameter with dimensions of length. The Chern-Simons coupling
constant is related to Newton's constant by 
\begin{equation}
k = \frac{l}{8G}.
\label{k}
\end{equation}
In order to agree with the conventions followed in \cite{BTZ}, we
use units in which $G=1/8$ and hence $k=l$. 

\subsection{Boundary conditions. A 1+1 generally covariant theory}

Consider the Chern-Simons action for the group $SO(2,1)\times
SO(2,1)$.  We apply the boundary condition (\ref{A0}) to each
$SO(2,1)$ copy, thus we impose
\begin{equation}
A^a_0 = w A^a_\varphi, \mbox{\hspace{1cm}} \tilde A^a_0 =
\tilde{w} \tilde
A^a_\varphi.
\label{A02}
\end{equation}
The boundary term coming from the variation of the Chern-Simons
action is
\begin{equation}
\frac{k}{4\pi}\int (\delta w A^2 - \delta\tilde{w}
\tilde A^2)  + \delta B.
\label{B22}
\end{equation}  
We shall shortly impose some conditions over the functions $w$
and $\tilde w$. However, it is convenient to keep them as
arbitrary functions in order to clarify their geometrical meaning.
If $w$ and $\tilde w$ are arbitrary functions of time, we
get at the boundary the two Virasoro constraint equations 
\begin{equation}
 L=(k/2)A^2=0,\mbox{\hspace{1cm}} \tilde L=(-k/2)\tilde A^2=0
\label{L2=0}
\end{equation}
ensuring the vanishing of the boundary term (\ref{B22}), with $B=0$.
It is a standard result that if $L$ and $\tilde L$ satisfy the
Virasoro algebra, then the combinations $H= L -\tilde L$ and
$H_\varphi = L+\tilde L$ satisfy the Dirac 1+1 deformation algebra.
This means that the induced theory at the boundary is diffeomorphism
invariant. $H$ represents the generator of timelike deformations
(conveniently densitized) and $H_\varphi$ is the generator of
diffeomorphisms along $\varphi$.  The induced theory is
then given by the 2 copies of the $SO(2,1)$ Kac-Moody currents
subject to the constraints equations (\ref{L2=0}) or, equivalently,
$H=0$ and $H_\varphi=0$. The boundary action can thus be written as  
\begin{equation}
I = I_{\mbox{\tiny WZW}}(U) - I_{\mbox{\tiny WZW}}(\tilde U) + \int
dt d\varphi (N^\perp H +
N^\varphi H_\varphi),
\label{I2}
\end{equation} 
with $N^\perp=(w-\tilde w)/2$ and
$N^\varphi=(w+\tilde w)/2$.  The theory described by the
action (\ref{I2}), which can be understood as a non-Abelian string
theory in six dimensions, is certainly interesting in its own right.
(Unitary representations for (one copy of) the above action have been
found in \cite{Bars}.) However, in our application to black hole
physics we shall make some simplifications and consider only a
special case. First, we shall impose that, at the horizon, the
lapse function $N^\perp$ vanishes,
\begin{equation}
N^\perp=0. 
\label{N=0}
\end{equation} 
This condition is quite natural for a black hole. Indeed, at the
horizon (in these coordinates) the lapse $N^\perp$ vanishes on-shell.   
Second, since $H_\varphi$  is the generator of
diffeomorphisms along $\varphi$,  $N^\varphi$ represents the angular
velocity of the horizon. We use the same condition as in the last
section, $\partial_{\varphi}N^\varphi=0$, and define 
\begin{equation}
N^\varphi \equiv w(t).
\label{w2}
\end{equation}
Under conditions (\ref{N=0}) and (\ref{w2}) not all the
equations (\ref{L2=0}) are imposed at the boundary. Actually, only
one of them is imposed, namely, the zero mode (total) Virasoro
operator
\begin{equation}
L_0+\tilde L_0 = \frac{k}{2}(A^2 - \tilde
A^2)|_{zero\ mode} =0.
\label{L02}
\end{equation}
Since $N^\perp$ is fixed by (\ref{N=0}) and the non-zero modes of
$N^\varphi$ are fixed by (\ref{w2}) the other modes of Eqs.
(\ref{L2=0}) are not imposed.  The boundary action appropriate to the
boundary conditions (\ref{w2}) and (\ref{N=0}) is a modification of
(\ref{I2}),
\begin{equation}
I = I_{\mbox{\tiny WZW}}(U) - I_{\mbox{\tiny WZW}}(\tilde U) +
\frac{1}{2\pi}\int dt d\varphi 
\ w(t) (L_0 +\tilde L_0).  
\label{I3}
\end{equation}
This is Carlip's boundary action and its quantization gives rise to
the 2+1 black hole entropy.     

Before going to the quantization of this action let us clarify some
of the differences between (\ref{I2}) and (\ref{I3}).  In (\ref{I2})
there are two constraints per point which are a consequence of the
arbitrariness of the Lagrange multipliers $N^\perp$ and $N^\varphi$.
In (\ref{I3}), on the other hand, the Lagrange multipliers are
severely restricted by (\ref{w2}) and (\ref{N=0}) hence there is only
one constraint, $L_0+\tilde L_0=0$. Furthermore, from (\ref{B22}) we
see that $N^\perp$ and $L_0-\tilde L_0$ are conjugate pairs (in a
radial quantization) thus, fixing $N^\perp=0$ imply that $L_0-\tilde
L_0$ is undetermined. This will have an important consequence in the
next section. 

\subsection{Quantization and counting of states}

The quantization of this system is implemented with the quantum
version of (\ref{L02}), 
\begin{equation}
(L_{0}+\tilde L_0)|\psi> = 0 
\label{qc}
\end{equation}
plus the condition that the eigenvalues of $L_0-\tilde L_0$ are
undetermined.    
 
The states of the theory are then defined by representations of the
two Kac-Moody algebras $\{ T^{a}_{n},\tilde T^{a}_{n}\}$ subject to
the constraint (\ref{qc}).  It is standard to consider only highest
weight representations which are determined by a representation of
the subalgebra $\{ T^a_0,\tilde T^a_0 \}$ [the two copies of
$SO(2,1)$] which acts as vacuum state, and the value of the central
charge $k$. We shall parameterize the Casimir operators $C$ and
$\tilde{C}$ in the form,  
\begin{eqnarray}
C &=& 2\eta_{ab}T^{a}_{0}T^{b}_{0}=2(r_{+}-r_{-})^2, \nonumber\\
\tilde{C} &=& 2\eta_{ab}\tilde{T}^{a}_{0}\tilde{T}^{b}_{0}= 2(r_{+}
+r_-)^2.  
\label{C2}
\end{eqnarray}
The parameters $r_+$ and $r_-$ can be identified, on shell, with the
outer and inner horizons of the black
hole solution \cite{BTZ}. They are also related with the
$SO(2,1)\times SO(2,1)$ holonomy existing in the black hole topology
\cite{holonomy}.  Of course, the area of the outer horizon is equal
to $2\pi r_+$.  Note that $r_+$ and $r_-$ are also related to the
mass and angular momentum of the solution through $M=(r_+^2 +
r_-^2)/l^2$ and $J =(2r_+ r_-)/l$. However, mass and angular momentum
are concepts defined at infinity while $r_+$ and $r_-$ depend only on
the topology.   

Using (\ref{C2}) and the normal ordered expression for the Virasoro
operators (\ref{qL01}), the constraint equation (\ref{qc})
reads\cite{Carlip},
\begin{equation}
L_{0}+\tilde L_0 = \hbar(N+\tilde N) +  Q^2
 - \frac{4r_+^2}{\hbar}=0
\label{qL0}
\end{equation}
and the combination $L_0-\tilde L_0$ reads
\begin{equation}
L_{0}-\tilde L_0 = \hbar (N-\tilde N) + \frac{\hbar Q^2}{2k} +
\frac{2r_-^2}{k}\equiv H.
\label{qH}
\end{equation}
Here $N$ and $\tilde N$ are number operators for each affine
$SO(2,1)$ algebra, and $Q^2$ is a shorthand for
\begin{equation}
Q^2=\frac{4\hbar}{4k^{2}-\hbar^2}
\left(\frac{2kr_{+}}{\hbar} - r_- \right)^2.
\label{Q}
\end{equation}
We showed at the end of last section that the
operator $H$ is canonically conjugate to the lapse function
$N^\perp$ and since $N^\perp$ at the horizon has been set equal to
zero, the eigenvalues of $H$ are undetermined. 

We now count states with a fixed value of $r_+$. In the limit in
which the number operators $N$ and $\tilde N$ are large the
difference $N-\tilde N$ approaches to zero. Since $r_+$
is fixed and $H$ is undetermined, Eq. (\ref{qH}) implies that $r_-$
is undetermined.  Eq. (\ref{qL0}), on the other hand, expresses the
number operator $N+\tilde N$ in terms of $r_{+}$ and $r_{-}$. Since
$r_-$ is undetermined, we have to sum over its possible values.  In
the thermodynamical limit, the largest contribution to the
number operator comes from  $Q=0$  ($r_{-} = 2kr_{+}/\hbar$)
and one obtains $N = \left(2r_+/\hbar\right)^2$. As shown
in \cite{Carlip} the logaritm of the degeneracy of states produces an
entropy given by 
\begin{equation}
S=\frac{2\pi r_{+}}{4\hbar}
\label{S}
\end{equation}
which coincides exactly with the Bekenstein-Hawking value for the 
2+1 black hole entropy. 

In this calculation there is one point that deserves special
attention. The boundary theory was defined for
any value of the level $k$. However, the calculation of the entropy
makes use of the limit $k \rightarrow \infty$. This limit is
necessary because the $SL(2,\Re)$ WZW model is not completely
understood (although unitary representations have been found in
\cite{Bars}). It is rather odd that at the very end we need to use
that limit since we do not know how to count states in the full
non-Abelian theory.  A striking feature of this calculation is the
fact that the non-Abelian nature of the theory does play a central
role anyway. Indeed, the result (\ref{S}) depends crucially in the
shift of the coupling constant $k\rightarrow k + \hbar q/2$ induced
by the non-Abelian Sugawara construction [see (\ref{qL01})]. Had
we taken the limit $k\rightarrow \infty$ at the very beginning, we
would not have obtained the right value for the black hole entropy
\cite{Carlip}.

\acknowledgments
During this work we have benefit from many conversations with Claudio
Teitelboim and Ricardo Troncoso. We also acknowledge useful
correspondence with Steve Carlip and Miguel Ortiz.  This work was
partially supported by grants \# 1960065, \# 1940203 and \# 3960008
from FONDECYT (Chile), and institutional support by a group of
Chilean companies (EMPRESAS CMPC, CGE, COPEC, CODELCO, MINERA LA
ESCONDIDA, NOVAGAS, ENERSIS, BUSINESS DESIGN ASS. and XEROX Chile).

\end{document}